\documentclass[fleqn,usenatbib]{mnras}

\usepackage[T1]{fontenc}

\DeclareRobustCommand{\VAN}[3]{#2}
\let\VANthebibliography\thebibliography
\def\thebibliography{\DeclareRobustCommand{\VAN}[3]{##3}\VANthebibliography}

\usepackage{graphicx}	% Including figure files
\usepackage{amsmath}	% Advanced maths commands
\usepackage{amssymb}	% Extra maths symbols
\usepackage{caption}
\usepackage{subcaption}
\usepackage{ulem}
\usepackage{mathrsfs}
\usepackage{newtxtext,newtxmath}
\title[Transient $\gamma$-ray QPOs in PKS 1510-089]{Transient Quasi-Periodic Oscillations at $\gamma$-rays in the TeV Blazar PKS 1510-089}

\author[Roy et al. 2021]{
Abhradeep Roy,$^{1}$\thanks{E-mail: abhradeep.roy@tifr.res.in}
Arkadipta Sarkar,$^{1}$
Anshu Chatterjee,$^{1}$
Alok C. Gupta,$^{2}$
Varsha Chitnis,$^{1}$
and P. J. Wiita$^{3}$
\\
$^{1}$Department of High Energy Physics, Tata Institute of Fundamental Research, Homi Bhabha Road, Mumbai-400005, India\\
$^{2}$Aryabhatta Research Institute of Observational Sciences (ARIES), Manora Peak, Nainital, 263002, India \\
$^{3}$Department of Physics, The College of New Jersey, PO Box 7718, Ewing, NJ 08628-0718, USA
}

\date{Accepted XXX. Received YYY; in original form ZZZ}

\pubyear{2021}

\begin{document}
\label{firstpage}
\pagerange{\pageref{firstpage}--\pageref{lastpage}}
\maketitle

\begin{abstract}
We present periodicity search analyses on the $\gamma$-ray lightcurve of the TeV blazar PKS 1510-089 observed by the \textit{Fermi} Large Area Telescope. We report the detection of two transient quasi-periodic oscillations: a 3.6-day QPO during the outburst in 2009 that lasted five cycles (MJD 54906--54923); and a periodicity of 92 days spanning over 650 days from 2018 to 2020 (MJD 58200--58850), which lasted for seven cycles. We employed the Lomb-Scargle periodogram, Weighted Wavelet Z-transform, REDFIT, and the Monte Carlo lightcurve simulation techniques to find any periodicity and the corresponding significance. The 3.6-day QPO was detected at a moderate significance of $\sim$3.5$\sigma$, while the detection significance of the 92-day QPO was $\sim$7.0$\sigma$.  We explore a few physical models for such transient QPOs including a binary black hole system, precession of the jet, a non-axisymmetric instability rotating around the central black hole near the innermost stable circular orbit, the presence of quasi-equidistant magnetic islands inside the jet, and a geometric model involving a plasma blob moving helically inside a curved jet. 
\end{abstract}


%%%%%%%%%%%%%%%%%%%%%%%%%%%%%%%%%%%%%%%%%%%%%%%%%%

%%%%%%%%%%%%%%%%% BODY OF PAPER %%%%%%%%%%%%%%%%%%

\section{Introduction}

%%%%%%%%%%%%%%%%%%%%%%%%%%%%%%%%%%%%%%%%%%%%%%%%%%%%%%%%%%%%%%%%%
\begin{figure*}
\centering
\begin{subfigure}{0.98\textwidth}
    \centering
    \includegraphics[width=\linewidth]{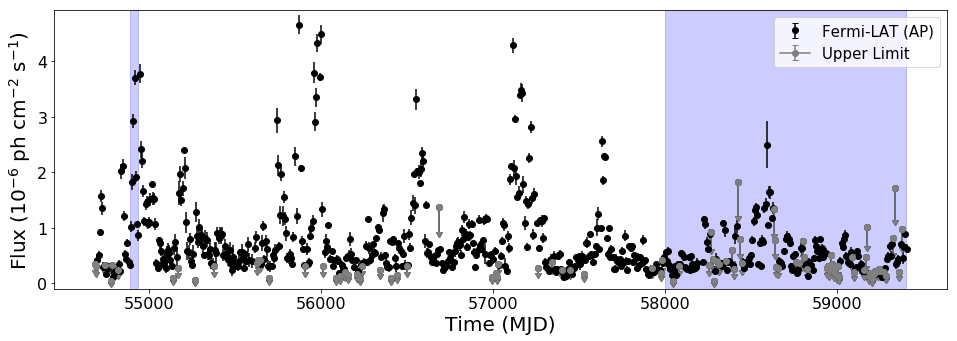}
    \caption{}
    \label{fig:fig1}
\end{subfigure}
\begin{subfigure}{0.98\textwidth}
    \centering
    \includegraphics[width=\linewidth]{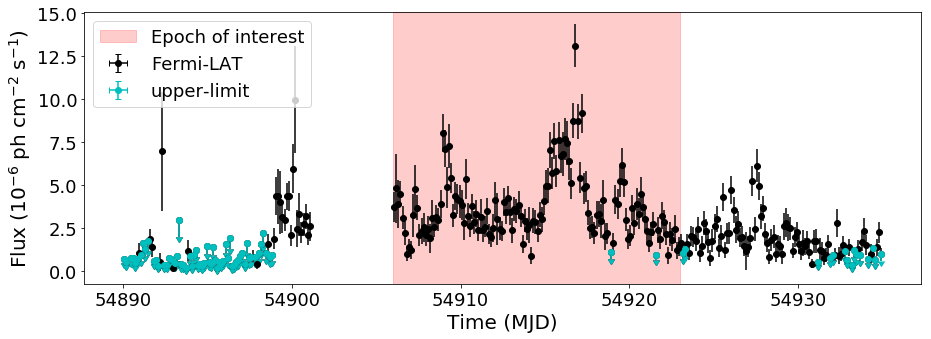}
    \caption{}
    \label{fig:fig2}
\end{subfigure}
\begin{subfigure}{0.98\textwidth}
    \centering
    \includegraphics[width=\linewidth]{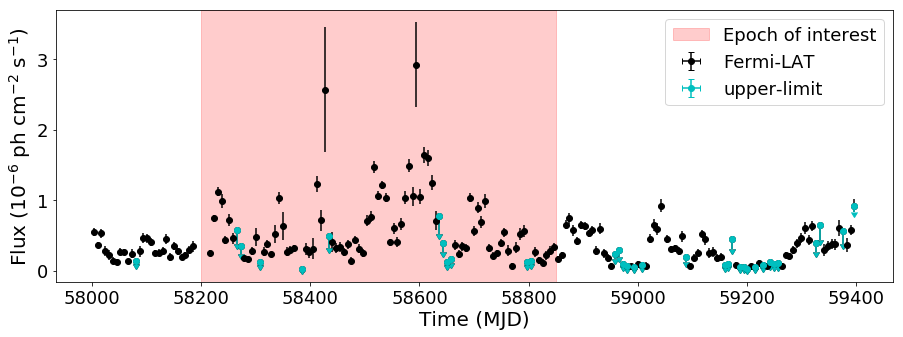}
    \caption{}
    \label{fig:fig3}
\end{subfigure}
\caption{\textbf{Top:} \textit{Fermi} 12-year weekly aperture photometry light curve. The blue shaded regions are the epochs where QPO analyses were carried out (i.e., MJD 54890--54935 and MJD 58000--59400). \textbf{Middle:} \textit{Fermi} 3-hour binned light curve within MJD 54890--54935. The pink shaded region is the epoch MJD 54906--54923 (EP1) where the final QPO analyses were carried out. \textbf{Bottom:} \textit{Fermi} 7-day binned light curve within MJD 58000--59400. The pink shaded region is the epoch MJD 58200--58850 (EP2) where the final QPO analyses were carried out.}
\label{fig:fig}
\end{figure*}
%%%%%%%%%%%%%%%%%%%%%%%%%%%%%%%%%%%%%%%%%%%%%%%%%%%%%%%%%%%%%%%%%%%%

Active Galactic Nuclei (AGN) are known to exhibit highly variable emission across the electromagnetic spectrum. Radio-loud AGN contain a pair of highly collimated relativistic plasma jets emanating from the central super-massive black hole (SMBH).  The jets are powered by the accretion process of dense ionized gases onto the SMBH. Radio-loud AGNs, with jets oriented close to our line of sight, form a subclass called the blazars \citep{Urry_1995}. Blazars are classified further into BL Lacertae objects (BL Lacs) and flat spectrum radio quasars (FSRQs) based on the strength of emission lines present in their optical-UV spectra. BL Lac spectra contain very weak and narrow emission lines, whereas FSRQs show broad and strong emission lines. The Doppler-boosted jet radiation dominates the blazar non-thermal emission from radio to very high energy (VHE) $\gamma$-ray wavebands. Various observational studies show that the blazars display flux variability of the order of minutes to years at $\gamma$-ray waveband, as observed by the \textit{Fermi}-LAT and ground based atmospheric Cherenkov telescopes \citep{Aleksi__2011, Shukla_2018}.

Although the nature of blazar variability is mostly non-linear, stochastic and aperiodic \citep{Kushwaha_2017}, many studies have claimed detections of strong quasi-periodic oscillations (QPO) in blazar lightcurves of different electromagnetic wavebands. In accordance to the variability timescales, the reported QPOs range from a few tens of minutes to hours to days and even years of timescales \citep[ and references therein]{2001A&A...377..396R, Liu_2006, 2009ApJ...690..216G, 2009A&A...506L..17L, 10.1093/mnrasl/slt125, 2014JApA...35..307G, 2018Galax...6....1G, 2019MNRAS.484.5785G, Ackermann_2015, 2018NatCo...9.4599Z, 10.1093/mnras/stz1482, 10.1093/mnras/staa3211,2020A&A...642A.129S}. However, most of the QPOs claimed in older studies are marginal detections in that they lasted for only a 2--4 cycles, while their significances were overestimated \citep{2014JApA...35..307G}.

Continuous monitoring of blazars by \textit{Fermi}-LAT in the last 12-years has led to a few recent highly significant $\gamma$-ray QPO detections in different blazars, including a $\sim$34.5-day transient QPO in PKS 2247-131 \citep{2018NatCo...9.4599Z}, a $\sim$71-day transient QPO in B2 1520+31 \citep{2019MNRAS.484.5785G}, a $\sim$47-day QPO in 3C 454.3 \citep{10.1093/mnras/staa3211}, a fast periodicity of $\sim$7.6 days in CTA 102 during an outburst \citep{2020A&A...642A.129S}, and a periodicity of $\sim$314 days in OJ 287 \citep{2020MNRAS.499..653K}. Although \citet{2019MNRAS.482.1270C} did the periodogram analysis on the \textit{Fermi}-LAT aperture photometry lightcurves of 10 blazars and claimed the absence of any global significant periodicity in $\gamma$-rays, a more systematic approach to $\gamma$-ray QPO detection by \citet{2020ApJ...896..134P}, involving multiple independent techniques applied on about 2300 AGNs, revealed the presence of global periodicities of $>4\sigma$ significance in 11 sources, along with 13 more sources with moderately significant (3--4$\sigma$) QPOs.

According to the leptonic models for jet dominated blazar emission, the source of the radio through optical-UV photons from blazars is the synchrotron emission by the dense population of ultra-relativistic electrons inside the magnetized jet.  External photon fields from the accretion disk, the broad-line region (BLR), and the dusty torus can enter the jet. Along with the synchrotron photons, these external photons get Compton-upscattered by the same relativistic electron population and produce the high energy $\gamma$-rays, often up to TeV energies. A few recent studies claimed probably related QPOs in optical and $\gamma$-rays for several sources, although the significances were low \citep{Sandrinelli_2016, Sandrinelli_2016a}. \citet{Ackermann_2015} reported a significant $\sim$2.2-year periodicity in highly-correlated optical and $\gamma$-ray light curves of PG 1553+113. Several recent studies have reported simultaneous oscillations in both optical and $\gamma$-ray light curves of the blazars BL Lac \citep{2017A&A...600A.132S}, 3C 454.3 \citep{10.1093/mnras/staa3211} and CTA 102 \citep{2020A&A...642A.129S}.

PKS 1510-089 (R.A. = 15h 12m 52.2s, Dec. = -09$^{\circ}$ 06' 21.6") belongs to the FSRQ subclass of blazars, and it is situated at a cosmological redshift of $z = 0.361$. It is one of the most well-studied blazars and shows high variability across all electromagnetic wavebands \citep{2018heas.confE..33Z}. The H.E.S.S. telescope detected PKS 1510-089 in 2010 at TeV energies \citep{2013A&A...554A.107H}. It exhibits occasional huge multi-wavelength outbursts crossing the daily $\gamma$-ray flux level of 10$^{-5}$ photons cm$^{-2}$ s$^{-1}$ \citep{2014A&A...567A.113B, Prince_2017, Meyer_2019}, as well as orphan $\gamma$-ray flaring episodes \citep{2021JHEAp..29...31P}. These correlated multi-wavelength flares can be explained using a shock-in-jet model that indicates the formation of an emission component in the compact core region, resulting in optical and $\gamma$-ray flares \citep{2017A&A...606A..87B}. Recently, PKS 1510-089 has been subject to several QPO studies at different wavelengths. \citet{Sandrinelli_2016} carried out QPO searches on about 3000-day long quasi-simultaneous multi-wavelength lightcurves (MJD 54000--57000) and reported modestly significant  periodicities of 115 days (<3$\sigma$) in $\gamma$-rays, 206 and 490 days ($\sim$3$\sigma$) in the optical-R band, and 207 and 474 days (>3$\sigma$) in the IR-K band. \citet{10.1111/j.1365-2966.2005.09150.x} detected periodic deep flux minima of around 1.84 years in its optical observations of past few years and inferred the presence of a binary black hole system at the centre. Based on 27-year long UMRAO data of PKS 1510-089 at 4.8, 8 and 14.5 GHz, \citet{2007A&A...462..547F} claimed a possible $\sim$12-year periodicity. The 15-year long 22 and 37 GHz radio data from 1990 to 2005 revealed two periodicities of 0.9 years and 1.8 years, which agrees well with the deep flux minima periodicity of 1.84 years and thereby, strengthens the binary black hole assumption \citep{2008AJ....135.2212X, 2009ScChG..52.1442Z, 2014AcASn..55....1F}. According to \citet{2017A&A...601A..30C}, QPO analysis of hard X-ray data of PKS 1510-089 taken by RXTE-PCA from 1996 to 2011 reveals no obvious periodicity. The most recent QPO study on radio data of PKS 1510-089 spanning over 38 years detects quite significant ($>4\sigma$) QPOs of $\sim$570, $\sim$800 and $\sim$1070 days in the 8 GHz and 14.5 GHz light curves \citep{2021JApA...42...92L}.

In this paper, we report the detection of a probable fast $\gamma$-ray periodicity of 3.6 days on top of the outburst in 2009 with $>$3.5$\sigma$ significance and a strong $\sim$3-month periodicity during a moderate state of activity between 2018 and 2020 with a significance above 7$\sigma$. We look for the plausible scenarios among several QPO models proposed in the literature to explain these transient QPOs. We consider that the 3.6-day QPO might have resulted from a hotspot rotating close to the innermost stable circular orbit around the central SMBH, or from enhanced emission from a few quasi-equidistant magnetic islands inside the jet. The source of the 92-day QPO could well be a helical motion of a plasma blob inside a curved jet. We state the \textit{Fermi}-LAT data analysis procedure in \autoref{sec:obs}, and describe the QPO finding algorithms in \autoref{sec:qpo}. Then we summarize our major results in \autoref{sec:res}. We try to interpret our key results on the basis of various physical models in \autoref{sec:disc} and summarize our conclusions in \autoref{sec:con}.

%%%%%%%%%%%%%%%%%%%%%%%%%%%%%%%%%%%%%%%%%%%%%%%%%%%%%%%%%%%%%%%%%%
\begin{figure*}
\centering
\begin{subfigure}{0.48\textwidth}
    \centering
    \includegraphics[width=\textwidth]{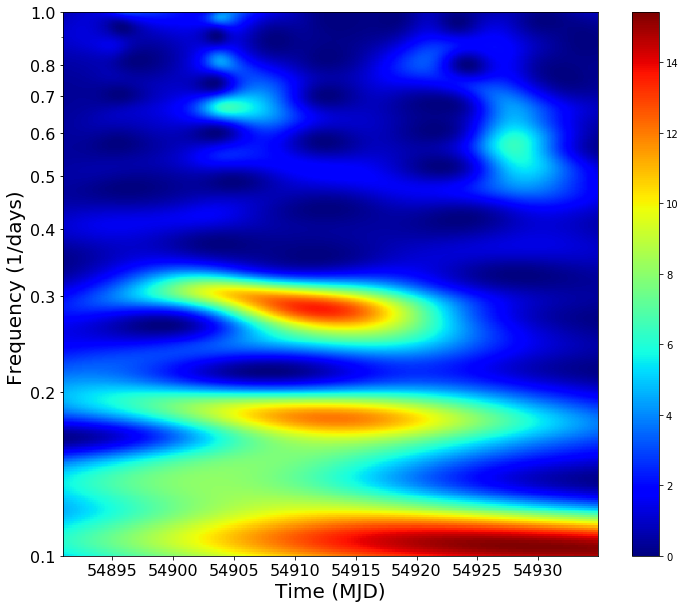}
    \caption{}
    \label{fig:wwz45}
\end{subfigure}
\begin{subfigure}{0.516\textwidth}
    \centering
    \includegraphics[width=\textwidth]{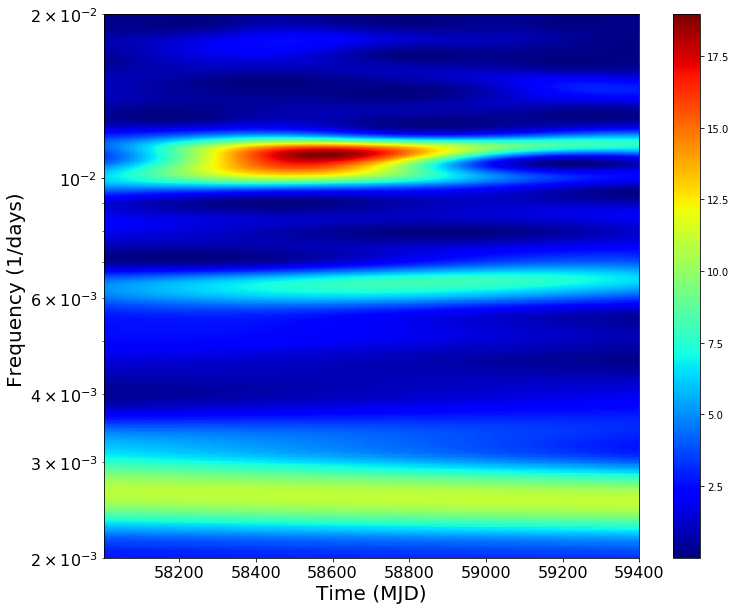}
    \caption{}
    \label{fig:wwz1400}
\end{subfigure}
\caption{\textbf{left:} WWZ map of PKS 1510-089 $\gamma$-ray light curve in the interval of MJD 54890–-54935. The bright yellowish red patch around the frequency of 0.3 days$^{-1}$ indicates the possible presence of a QPO in the interval of MJD 54900–-54923. \textbf{right:} WWZ map of PKS 1510-089 $\gamma$-ray lightcurve in the interval of MJD 58000--59400. The bright red patch indicates a probable QPO in the interval of MJD 58200--58850.}
\label{fig:wwz_prelim}
\end{figure*}
%%%%%%%%%%%%%%%%%%%%%%%%%%%%%%%%%%%%%%%%%%%%%%%%%%%%%%%%%%%%%%%%%%%

%%%%%%%%%%%%%%%%%%%%%%%%%%%%%%%%%%%%%%%%%%%%%%%%%%%%%%%%%%%%%%%%%%%
\begin{figure*}
    \centering
\begin{subfigure}{0.48\textwidth}
    \centering
    \includegraphics[width=\textwidth]{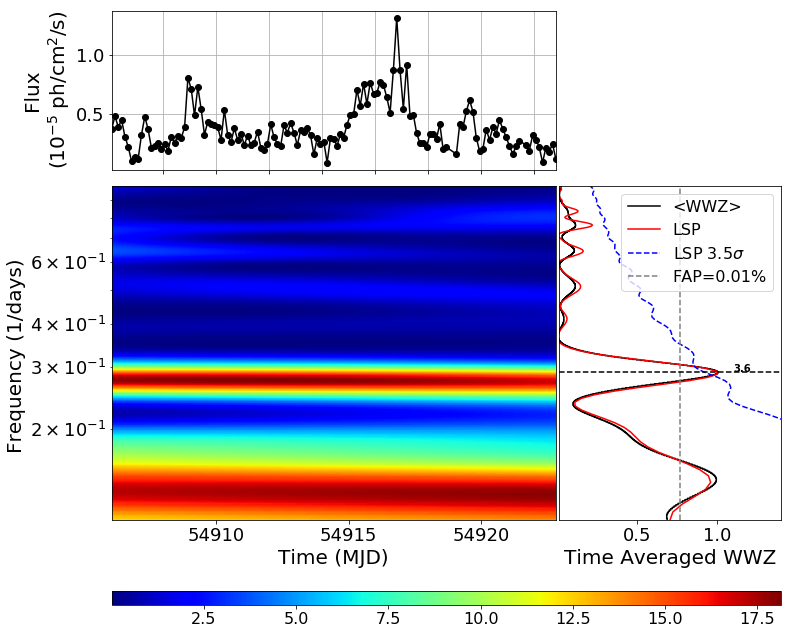}
    \caption{}
    \label{fig:wwzep1}
\end{subfigure}
\begin{subfigure}{0.48\textwidth}
    \centering
    \includegraphics[width=\textwidth]{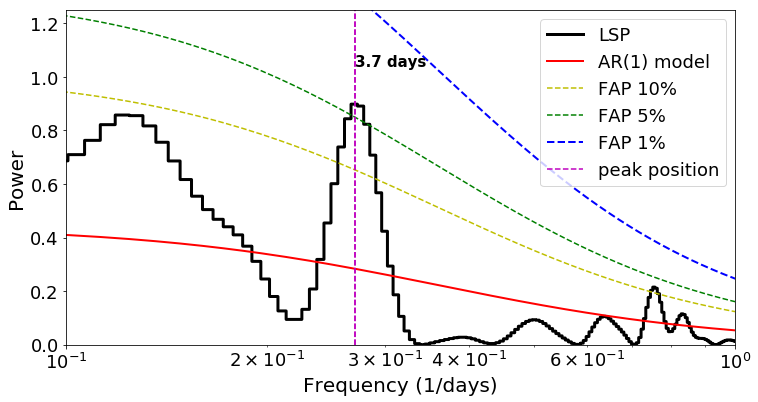}
    \caption{}
    \label{fig:rfep1}
\end{subfigure}
\begin{subfigure}{0.48\textwidth}
    \centering
    \includegraphics[width=\textwidth]{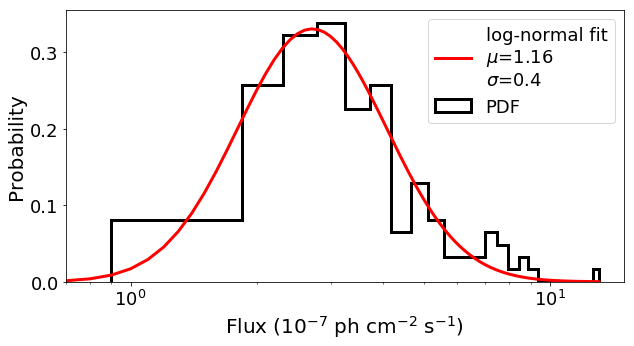}
    \caption{}
    \label{fig:pdfep1}
\end{subfigure}
\begin{subfigure}{0.48\textwidth}
    \centering
    \includegraphics[width=\textwidth]{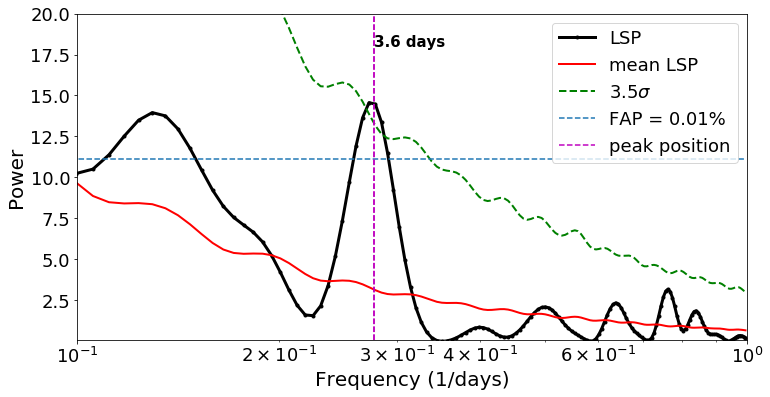}
    \caption{}
    \label{fig:lcsimep1}
\end{subfigure}
    \caption{\textit{QPO analysis in EP1 (MJD 54906--54923):} \textbf{Upper-left:} WWZ map of PKS 1510-089 $\gamma$-ray light curve during EP1. On the upper-left sub-panel, the light curve is shown; the lower left panel shows the WWZ map; the lower-right sub-panel shows the time-averaged WWZ (black) as well as the LSP (red). The blue dashed line represents the 3.5$\sigma$ significance line against the power-law red-noise spectrum and the red band on the WWZ map indicates a strong periodicity of $\sim$3.6 days. \textbf{Upper-right:} Power-spectrum of PKS 1510-089 $\gamma$-ray light curve using \texttt{REDFIT} during EP1. The black line represents the power-spectrum, the red line is the theoretical AR1 spectrum, and the yellow, green and blue dashed lines represent FAP levels of 10\%, 5\% and 1\%, respectively. A strong periodicity of $\sim$3.7 days crosses the 5\% FAP level. \textbf{Lower-left} Flux distribution of the PKS 1510-089 $\gamma$-ray light curve (PDF in black) fitted with a log-normal model (red) that is used as an input in light curve simulation. \textbf{Lower-right} Result of light curve simulation of the PKS 1510-089 $\gamma$-ray light curve during EP1. The black line represents the LSP of the original light curve and the red line is the mean LSP of the simulated light curves. The dominant period of $\sim$3.6 days crosses the 0.01\% FAP level (cyan dashed line) and the 3.5$\sigma$ significance curve (green dashed curve).}
    \label{fig:qpoep1}
\end{figure*}
%%%%%%%%%%%%%%%%%%%%%%%%%%%%%%%%%%%%%%%%%%%%%%%%%%%%%%%%%%%%%%%%%%%

%%%%%%%%%%%%%%%%%%%%%%%%%%%%%%%%%%%%%%%%%%%%%%%%%%%%%%%%%%%%%%%%%%%
\begin{figure*}
    \centering
\begin{subfigure}{0.48\textwidth}
    \centering
    \includegraphics[width=\textwidth]{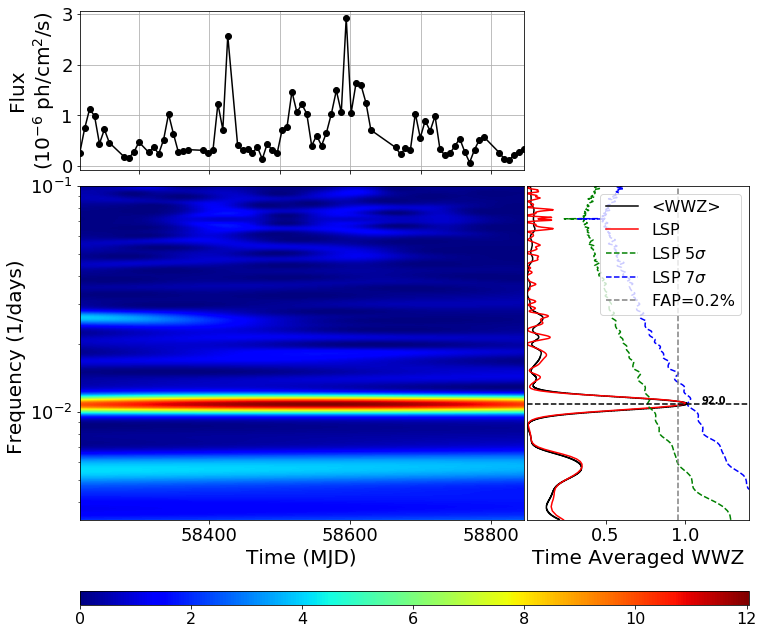}
    \caption{}
    \label{fig:wwzep2}
\end{subfigure}
\begin{subfigure}{0.48\textwidth}
    \centering
    \includegraphics[width=\textwidth]{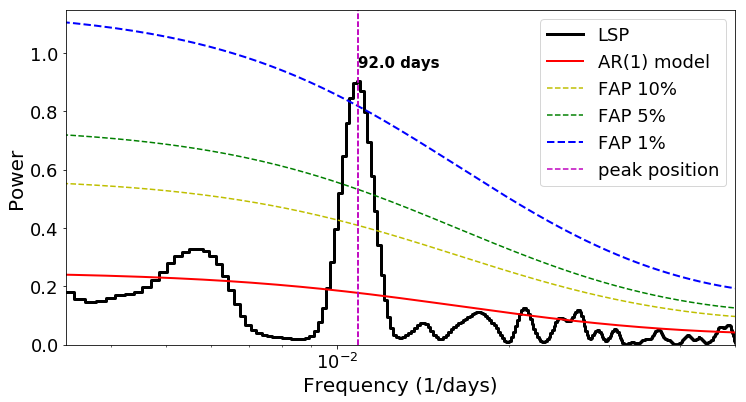}
    \caption{}
    \label{fig:rfep2}
\end{subfigure}
\begin{subfigure}{0.48\textwidth}
    \centering
    \includegraphics[width=\textwidth]{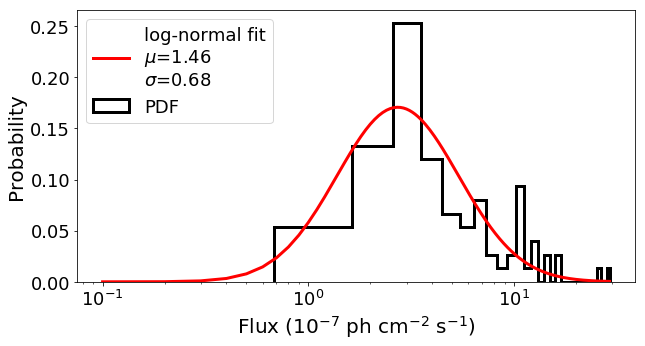}
    \caption{}
    \label{fig:pdfep2}
\end{subfigure}
\begin{subfigure}{0.48\textwidth}
    \centering
    \includegraphics[width=\textwidth]{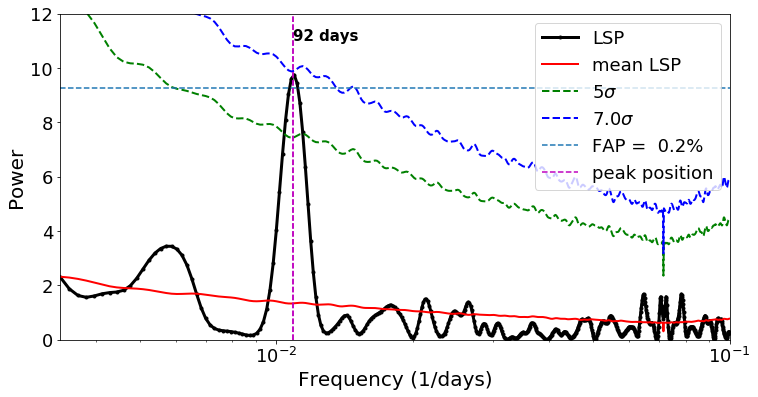}
    \caption{}
    \label{fig:lcsimep2}
\end{subfigure}
    \caption{\textit{QPO analysis in EP2 (MJD 58200--58850):} \textbf{Upper-left:} WWZ map of the PKS 1510-089 $\gamma$-ray light curve during EP2. On the upper-left sub-panel, the light curve is shown, the lower left sub-panel displays the WWZ map and the lower-right sub-panel shows the time-averaged WWZ power (black) on top of the LSP (red). The green and blue dashed curves represent the 5.0$\sigma$ and 7.0$\sigma$ significance respectively against the power-law red-noise spectrum and the red band on the WWZ map indicates a strong periodicity of $\sim$92 days. \textbf{Upper-right} Power-spectrum of PKS 1510-089 $\gamma$-ray light curve using \texttt{REDFIT} during EP2. The black line represents the power-spectrum, the red line is the theoretical AR1 spectrum, and the yellow, green, and blue dashed lines represent FAP level of 10\%, 5\%, and 1\%, respectively. A strong periodicity of $\sim$92 days crosses the 1\% FAP level. \textbf{Lower-left:} PDF of the flux distribution of PKS 1510-089 $\gamma$-ray light curve (black)  fitted with a log-normal model (red) that is used as an input in the light curve simulations. \textbf{Lower-right:} Result of light curve simulations of PKS 1510-089 $\gamma$-ray light curve during EP2. The black line represents the LSP of the original light curve, and the red line is the mean LSP of the simulated light curves. The dominant period of $\sim$92 days crosses the 0.2\% FAP level (cyan dashed line) and the 5$\sigma$ significance level (green dashed curve) and nearly touches the  a significance of 7.0$\sigma$ (blue dashed curve).}
    \label{fig:qpoep2}
\end{figure*}
%%%%%%%%%%%%%%%%%%%%%%%%%%%%%%%%%%%%%%%%%%%%%%%%%%%%%%%%%%%%%%%%%%%

\section{Observations and Data Analysis}
\label{sec:obs}

A detailed recent study on $\gamma$-ray periodicity in AGN by \citet{2020ApJ...896..134P} showed that any long-lived QPO in PKS 1510-089 is not significant (<3$\sigma$). But on visual inspection, the weekly \textit{Fermi}-LAT aperture photometry light curve from MJD 58000 to MJD 59400 seemed to have some periodicity during a long moderate activity state. Moreover, the flare state light curve modelling by \citet{Prince_2017} indicated the presence of fast periodicity ($\sim$ days) during the flare between MJD 54890 to MJD 54935.

\subsection{\textit{Fermi}-LAT Data}
\label{sec:data} % used for referring to this section from elsewhere

We obtained the $\gamma$-ray data from the Large Area Telescope (LAT) facility, on-board the \textit{Fermi} observatory. The \textit{Fermi}-LAT is an imaging space-based telescope that detects $\gamma$-rays using the pair-production technique within the 30 MeV to 1 TeV energy range. LAT has a large angular field of view of about 2.3 sr and covers the entire sky every three hours \citep{2009ApJ...697.1071A}. We collected the PASS8 (P8R3) processed events' data of PKS 1510-089 between MJD 54890--54935 and MJD 58000--59400 from the \textit{Fermi}-LAT data archive \footnote{\url{https://fermi.gsfc.nasa.gov/ssc/data/access/}}. The PASS8 data provides a significant improvement in the data quality using an improved reconstruction of the entire LAT events \citep{Abdollahi_2020}.

\subsection{Data Reduction}
\label{sec:ana}

We used the standard software package \texttt{FERMITOOLS-v2.0.8} recommended by the \textit{Fermi}-LAT collaboration \citep{2019ascl.soft05011F} and the user-contributed python script \texttt{ENRICO} \citep{sanchez2013enrico}. Following the recommendations of the \textit{Fermi}-LAT collaboration\footnote{\url{https://fermi.gsfc.nasa.gov/ssc/data/analysis/}}, we chose the events belonging to the SOURCE class \texttt{(evclass=128, evtype=3)} within the energy range of 0.1--300 GeV from a circular region of interest (ROI) having a radius of 15$^{\circ}$ centred at the source PKS 1510-089. To get rid of the $\gamma$-ray contribution from the Earth's albedo, we selected the events having zenith angle less than 95$^{\circ}$ followed by the good time interval (GTI) selection using the standard filter \texttt{"(DATA$\_$QUAL$>0$)$\&\&$(LAT$\_$CONFIG==1)"}. We generated an XML file containing the spectral shapes of all the sources lying within ROI+10$^{\circ}$ radius around the source location according to the fourth \textit{Fermi}-LAT (4FGL) catalogue, including the $\gamma$-ray background emission templates \texttt{"gll\_iem\_v07.fits"} and \texttt{"iso\_P8R3\_SOURCE\_V3\_v1.txt"} for the Galactic and extra-galactic contributions respectively. We carried out an unbinned maximum likelihood analysis over the input XML spectral file using the \texttt{GTLIKE} tool to obtain the source spectrum using the instrumental response function \texttt{P8R3\_SOURCE\_V3}. Except for scaling factors, We kept all the spectral parameters free to vary during the optimization process for the sources lying within 5$^{\circ}$ from PKS 1510-089. The iterative likelihood analysis removed the sources having significance less than 1$\sigma$ after each fitting pass. As mentioned in the 4FGL catalogue, the final source spectrum was modelled using a log-parabola given as,
\begin{equation}
    \frac{dN}{dE} = k \left(\frac{E}{E_b}\right)^{-\alpha-\beta\log(E/E_b)},
\end{equation}
where $\alpha$ is the spectral index at the break energy ($E_b$). We kept $E_b$ fixed during the likelihood fitting process.

The full 12-year \textit{Fermi} unfiltered aperture photometry light curve (\autoref{fig:fig1}) was obtained from the \textit{Fermi} monitoring source list webpage\footnote{\url{https://fermi.gsfc.nasa.gov/ssc/data/access/lat/msl\_lc}}. To extract the $\gamma$-ray light curve, we divided the whole epoch in a number of time-bins of our required size and carried out the entire above-mentioned procedure in each time-bin. For the epoch of MJD 54890--54935, we made a light curve with 3-hour long bins (\autoref{fig:fig2}), and for the epoch of MJD 58000--59400, we set the bin size to 7 days (\autoref{fig:fig3}). For the time-bins where the test-statistics of the flux estimation were less than 16 (i.e., essentially $<4\sigma$ detection significance), we estimated flux upper limits at 95\% confidence level using the profile-likelihood method.

\section{QPO Analysis Methods}
\label{sec:qpo}

On visual inspection, the light curves indicated possible quasi-periodic modulations. To estimate the time-period of the modulation and the corresponding significance, we applied four different methods to analyse the light curves: the Generalized Lomb-Scargle periodogram (GLSP), Weighted Wavelet Z-transform (WWZ), REDFIT, and light curve simulations. These methods follow different approaches to detect periodicities and their corresponding significances in unevenly sampled time-series. Although we have used evenly binned \textit{Fermi}-LAT light curves, they became unevenly sampled due to the omission of flux upper-limits in the analysis processes. The details of these methods are discussed below.

\subsection{Generalized Lomb-Scargle Periodogram}

The periodogram is one of the most common methods to find periodicities in light curves, and it gives the power of flux modulations at different frequencies. For an evenly sampled light curve, the square of the modulus of its discrete Fourier transform gives the periodogram. But for irregular sampling, the Lomb-Scargle periodogram (LSP) method iteratively fits sinusoids with different frequencies to the light curve and constructs a periodogram from the goodness of the fit \citep{1976Ap&SS..39..447L, 1982ApJ...263..835S}. In this work, we used the Generalized LSP sub-package of the \texttt{PYASTRONOMY} python package\footnote{\url{https://github.com/sczesla/PyAstronomy}} \citep{pya}. Unlike the classical LSP, the GLSP fits a sinusoid plus a constant to the light curve and takes the errors associated to the measured fluxes into account \citep{2009A&A...496..577Z}. This code also provides the significance of a peak in the periodogram in terms of the False Alarm Probability (FAP) given as,
\begin{equation}
    FAP(Pn) = 1 - (1-prob(P > Pn))^M ,
\end{equation}
where the $FAP$ denotes the probability that at least one out of $M$ independent power values in a given frequency band of a white-noise periodogram is larger than or equal to the power threshold, $Pn$. In this work, a peak in a periodogram was considered to be significant when it crossed the 1\% FAP line. The peak position and its corresponding uncertainty were estimated by fitting a Gaussian to the dominant periodogram-peak. GLSP is an effective tool to find persistent periodicities. But it cannot usually detect transient periodicities, as the non-periodic part of the light curve decreases the goodness of GLSP sinusoid fit. Therefore, the power of the transient periodicity is reduced.

\subsection{Weighted Wavelet Z-transform}

We used the Weighted Wavelet Z-transform method to detect transient quasi-periodicities \citep{1996AJ....112.1709F}. The WWZ method convolves a light curve with a time- and frequency-dependent kernel and decomposes the data into time and frequency domains to create a WWZ map. In this work, we used the Morlet kernel \citep{doi:10.1137/0515056} having the following functional form,
\begin{equation}
    f[\omega(t-\tau)] = \exp[\iota\omega(t-\tau) - c\omega^2(t-\tau)^2] .
\end{equation}
Then the WWZ map is given as,
\begin{equation}
    W[\omega,\tau; x(t)] = \omega^{1/2} \int x(t)f^{\ast}[\omega(t-\tau)]\text{d}t ,
\end{equation}
where $f^{\ast}$ is the complex conjugate of the Morlet kernel $f$, $\omega$ is the frequency, and $\tau$ is the time-shift. This kernel acts as a windowed discrete Fourier transform having a frequency dependant window size of $\exp[- c\omega^2(t-\tau)^2]$. The WWZ map has the advantage of being able to detect both any dominant periodicities and the time spans of their persistence.

\subsection{REDFIT}

The \texttt{REDFIT}\footnote{\url{https://www.manfredmudelsee.com/soft/redfit/index.htm}} software calculates the bias-corrected power-spectrum of a time-series and provides the significance of the peaks in the spectrum \citep{SCHULZ2002421}. \texttt{REDFIT} fits the light curve with a first-order Autoregressive process (AR1) to estimate the underlying red-noise spectrum, which is the characteristic of variable $\gamma$-ray emission from blazars \citep{10.1093/mnras/sty2720}. The Autoregressive (AR) models assume that the present observation in a time-series is related to the past observations. Thus, large fluctuations in the light curve become less likely \citep{RePEc:eee:spapps:v:6:y:1977:i:1:p:9-24}. A discrete AR1 process $\mathscr{F}$ for times $t_i$ ($i=1,2,...,N$) with uneven spacing is given as,
\begin{equation}
\begin{aligned}
    \mathscr{F}(t_i) = \theta_i \mathscr{F}(t_{i-1}) + \epsilon(t_i), \\ \theta_i = \exp((t_{i-1}-t_i)/\tau) ,
\end{aligned}
\end{equation}
where $\tau$ is the characteristic timescale and $\epsilon$ denotes white-noise with zero mean. The power-spectrum of an AR1 model has the following analytical form \citep{percival_walden_1993},
\begin{equation}
    G_{rr}(f_j) = G_0 \frac{1-\theta^2}{1-2\theta\cos(\upi f_j / f_{Nyq}) + \theta^2} ,
\end{equation}
where $f_j$ denotes the discrete frequency up to the Nyquist frequency ($f_{Nyq}$) and $G_0$ is the average spectral amplitude. The “average autoregression coefficient” ($\theta$) is related to the arithmetic mean of the sampling intervals $\Delta t = (t_N-t_1)/(N-1)$ as, $\theta = \exp(-\Delta t/\tau)$, while the $\tau$ comes from the Welch-overlapped-segment-averaging \citep[WOSA,][]{1161901} of the LSP.

\texttt{REDFIT} estimates the significance of the peaks in the power-spectrum using FAP level up to the minimum of 1\%. 

\subsection{Lightcurve Simulation}

Another way to estimate the peak significance in a periodogram is to simulate light curves using a Monte Carlo method, following the power spectral density (PSD) and the flux distribution (PDF) of the original light curve \citep{2013MNRAS.433..907E}. Simple power-laws ($P \propto \nu^{-\alpha}$) give reasonably good approximations of the underlying red-noise PSDs of blazar lightcurves \citep{2005A&A...431..391V}. Hence, we used a power-law and a log-normal model to respectively fit the PSD and PDF of the original light curve. The log-normal model for PDF has the form,
\begin{equation}
    PDF(x) = \frac{1}{x \sigma \sqrt{2\uppi}} \exp \left[ -\frac{(\ln x - \mu)^2}{2 \sigma^2} \right] .
\end{equation}
Then we simulated 2000 light curves with the fitted PSD and PDF models as inputs using the \texttt{DELightcurveSimulation}\footnote{\url{https://github.com/samconnolly/DELightcurveSimulation}} code \citep{2016ascl.soft02012C}. The mean and standard deviation of the simulated light curve GLSP at each frequency allowed us to estimate the significance of the dominant periodicities.

\section{Results}
\label{sec:res}

\autoref{fig:fig1} shows the epochs on the weekly \textit{Fermi}-LAT full aperture photometry lightcurve that were initially considered for QPO analysis. But after analysing the \textit{Fermi}-LAT data, we selected two sub-intervals (\autoref{fig:fig2} and \autoref{fig:fig3}) to carry on detailed QPO analysis: MJD 54906--54923 (EP1) and MJD 58200--58850 (EP2).

\subsection{EP1: 16 March 2009 -- 2 April 2009 (MJD 54906--54923)}

From \autoref{fig:wwz45} it is evident that there is a QPO in MJD 54900--54923 around the frequency of 0.3 days$^{-1}$. But in \autoref{fig:fig2}, it can be clearly seen that there is a big gap in the data and in addition, the available flux points are mostly upper-limits during MJD 54890--54906. So, we avoided these intervals and carried out our QPO analyses in EP1 (MJD 54906--54923).

%\begin{figure}
%    \centering
%    \includegraphics[width=0.48\textwidth]{fig4_LSP_3hr.png}
%    \caption{LSP of PKS 1510-089 $\gamma$-ray lightcurve during EP1 (MJD 54906--54923). The blue line corresponds to the LSP. The cyan, orange and green line represent FAP levels of 10\%, 5\% and 1\% respectively. A periodicity of $\sim$3.6 days is evident here.}
%    \label{fig:lspep1}
%\end{figure}

\autoref{fig:wwzep1} shows the WWZ map of PKS 1510-089 during EP1. The strong horizontal red patch denoting the 3.6-day QPO spans the entire 18-day EP1 light curve, indicating the presence of 5 cycles. The time-averaged WWZ plot indicates a QPO of 3.6 days with 3.5$\sigma$ significance. \autoref{fig:rfep1} shows the power-spectrum and the corresponding peak significance obtained using \texttt{REDFIT}. It indicates a QPO of $\sim$3.7 days, crossing the 5\% FAP level. \autoref{fig:pdfep1} represents the log-normal fitted flux distribution in this epoch. \autoref{fig:lcsimep1} shows the LSP of PKS 1510-089 $\gamma$-ray light curve during EP1 and the result of significance check using our light curve simulation. The dominant periodicity of 3.63$^{+0.07}_{-0.07}$ days crosses the 0.01\% FAP level. We note that this FAP significance estimation only holds for Gaussian random noise, whereas blazars show red-noise type variability.  However, the light curve simulation, which does not rely on that noise assumption, indicates that the QPO of 3.6 days has a significance of 3.5$\sigma$.

In recent studies \citep{10.1093/mnras/staa2899, 10.1093/mnras/staa3211, 2020ApJ...896..134P,  2020A&A...642A.129S}, the reported blazar QPOs generally have $>$3$\sigma$ significance and cross the \texttt{REDFIT} FAP level of 1\%. In this case, the QPO spans only  18 days and the apparent periodicity is very fast ($\sim$3.6 days) which should be the reason for low significance in the \texttt{REDFIT} output. But from the light curve simulation, we find a $\sim$3.5$\sigma$ significance, which makes the QPO sufficiently significant to be reported. The QPO peak significances remained the same when we applied the same analysis procedures on light curves in EP1 with different bin-sizes, such as 4 hours or 5 hours.

\subsection{EP2: 23 March 2018 -- 2 January 2020 (MJD 58200--58850)}
%%%%%%%%%%%%%%%%%%%%%%%%%%%%%%%%%%%%%%%%%%%%%%%%%%%%%%%%%%%%%%%%%%%

%%%%%%%%%%%%%%%%%%%%%%%%%%%%%%%%%%%%%%%%%%%%%%%%%%%%%%%%%%%%%%%%%%%

\autoref{fig:fig3} shows the 7-day binned \textit{Fermi}-LAT lightcurve between MJD 58000 to MJD 59400. Although the GLSP did not reveal any significant dominant QPO in this interval, the WWZ map showed a bright red patch with a span of $\sim$650 days from MJD 58200 to MJD 58850 (\autoref{fig:wwz1400}). This led to the selection of EP2 (MJD 58200--58850) for detailed analysis.

%\begin{figure}
%    \centering
%    \includegraphics[width=0.48\textwidth]{fig10_LSP_58200_58850.png}
%    \caption{LSP of PKS 1510-089 $\gamma$-ray lightcurve during EP2 (MJD 58200--58850). The blue line corresponds to the LSP. The cyan, orange and green line represent FAP levels of 10\%, 5\% and 1\% respectively. A periodicity of $\sim$92 days is evident here.}
%    \label{fig:lspep2}
%\end{figure}

\autoref{fig:wwzep2} shows the WWZ map of PKS 1510-089 during EP2. The strong horizontal red patch, spanning all the 650-day EP2 light curve, denotes a strong 92-day QPO, indicating the presence of 7 cycles. The time-averaged WWZ also show a QPO of 92 days, with almost 7.0$\sigma$ significance. \autoref{fig:rfep2} shows the power-spectrum and the corresponding peak significance obtained using \texttt{REDFIT}. It indicates a QPO of $\sim$92 days,  crossing the 1\% FAP level. \autoref{fig:pdfep2} represent the log-normal fitted flux distribution during EP2. \autoref{fig:lcsimep2} shows the LSP of PKS 1510-089 $\gamma$-ray light curve during EP2 and the significance check of the observed QPO using light curve simulation. The dominant periodicity of 91.5$^{+1.2}_{-1.2}$ days crosses the 0.2\% FAP level, if one considers an underlying Gaussian-type noise. Light curve simulation shows that the QPO of 92 days seems to have a significance of $>6\sigma$. The 92-day peak touches the blue dashed 7.0$\sigma$ significance line. Thus, all the applied tools indicate this 92-day QPO to be highly significant. We have also checked that our key results remain the same when light curves produced when we employed different bin-sizes of 3 days or 5 days during EP2.

\section{Discussion:}
\label{sec:disc}
We obtained $\gamma$-ray data from the \textit{Fermi}-LAT archive and analysed then to generate the light curves in two different epochs: EP1 (MJD 54906--54923) and EP2 (MJD 58200--58850). We employed LSP, WWZ and \texttt{REDFIT} methods to detect significant transient periodicities in the light curves. We generated 2000 light curves in each epoch using Monte-Carlo simulation to account for the underlying red-noise spectrum while estimating the QPO significances. All of these methods revealed two highly probable transient QPOs: (1) a fast QPO of 3.6 days in EP1 and; (2) a 92-day QPO in EP2.

Due to lack of good coverage of optical observations during the selected epochs, it was not possible to employ QPO analysis tools on the available optical data. In case of an FSRQ, both synchrotron emission from the jet and thermal emission from the accretion disk contribute to the optical-UV emission. The $\gamma$-ray emission is dominated by the inverse-Compton scattering of seed photons by the charged particle population inside the jet responsible for the synchrotron emission. Thus, simultaneous optical and $\gamma$-ray QPO can lead to strong inferences about possible reasons behind such phenomena and the underlying disk-jet connection. Only having $\gamma$-ray data here, it is quite hard to conclusively investigate the probable physical reasons behind such periodicities. Still, the $\gamma$-ray QPO timescale and persistence can lead to a few insights about the jet structure or emission processes.

Earlier studies have proposed several possible models, such as supermassive binary black-hole systems \citep{2008Natur.452..851V, 10.1111/j.1365-2966.2009.16133.x, Ackermann_2015, 2021JApA...42...92L}, persistent jet precession model \citep{2000A&A...360...57R, Rieger_2004, 2018MNRAS.474L..81L}, and Lense-Thirring precession of accretion disks \citep{1998ApJ...492L..59S, 2018MNRAS.474L..81L} to explain the long-term quasi-periodicities in different blazar's emissions. Although PKS 1510-089 is a candidate to contain a binary black-hole system at the centre \citep{10.1111/j.1365-2966.2005.09150.x, 2007ChPhy..16..876L}, all these models exhibit persistent QPOs with at least year-long periods. So, we can probably discard these models as an explanation for the QPOs discussed here.  

Optical QPOs with tens of days period can be explained by hot-spots rotating at or near the innermost stable circular orbit (ISCO) around the central SMBH \citep{1991A&A...246...21Z, 2009ApJ...690..216G}. In this case, the rotation of the hot-spot will modulate the seed photon field of the external inverse-Compton scattering inside the jet, causing a modulation in the $\gamma$-ray emission. The $\gamma$-ray emission from blazar jet is Doppler boosted and should have a faster quasi-periodicity ($\sim$days) in this model. Thus, rotation of a hot-spot near the ISCO could be a possible scenario behind the $\sim$3.6-day QPO on top of the flare during EP1. Availability of an optical counterpart of this short QPO could actually emphasize the applicability of this model in EP1. Assuming that the QPO is related to orbital rotation of a hotspot, presence of a spiral shock or any other non-axisymmetric instabilities close to the ISCO, we can estimate the central SMBH mass using an expression given by \citet{2009ApJ...690..216G},
\begin{equation}
\label{eq:eq8}
    \frac{M_{\text{BH}}}{M_{\sun}} = \frac{3.23 \times 10^4 P}{(r^{3/2}+a)(1+z)} ,
\end{equation}
where $P$ is the QPO period in seconds, $z$ is the cosmological redshift of the source ($z$ = 0.361 for PKS 1510-089), $r$ is the radius of ISCO in units of $GM_{\text{BH}}/c^2$, and $a$  is the SMBH spin parameter.

For a Schwarzschild black hole, $r = 6.0$ and $a = 0$, and for a maximally rotating Kerr black hole, $r = 1.2$ and $a = 0.9982$. Using the period of $\sim3.6$ days, the estimated mass of a Schwarzschild BH is $\sim$5.0$\times10^8 M_{\sun}$, and the mass of a maximally rotating Kerr BH is $\sim$3.2$\times10^9 M_{\sun}$. While using the period of 92 days, we get the mass estimate of Schwarzschild BH to be $\sim$1.2$\times10^{10} M_{\sun}$, and that of a Kerr BH to be $\sim$8.2$\times10^{10} M_{\sun}$. There are various methods to estimate the mass of the central SMBH. \citet{10.1046/j.1365-8711.2001.04795.x} used the data of the H$_\beta$ line width and the optical continuum luminosity and reported a mass of $\sim$1.3$\times10^9 M_{\sun}$ for the central SMBH of PKS 1510-089. The reverberation mapping technique is one of the most accurate methods to estimate the mass of a primary black hole. \citet{2020A&A...642A..59R} reported a mass of $\sim$5.71$\times$10$^7 M_{\sun}$ for the SMBH of PKS 1510-089 using a stereoscopic reverberation mapping technique. \citet{2005AJ....130.2506X} used reverberation mapping and short-timescale optical variability to estimate the SMBH mass to be $\sim$2.0$\times10^8 M_{\sun}$ and $\sim$1.6$\times10^8 M_{\sun}$, respectively. The masses estimated using the 92-day QPO are much higher, thereby essentially excluding the possibility that disk instabilities directly yield it. But the masses estimated using the 3.6-day QPO are comparable to the SMBH mass estimated by \citet{10.1046/j.1365-8711.2001.04795.x}. If this is the case, it favours the presence of a non-maximally rotating SMBH at the centre.  However, this rotating hot-spot scenario suffers a major disadvantage in explaining blazar fluctuations. Blazar disks have almost face-on orientation with respect to the observer, so the motion of the rotating hotspot should be almost azimuthally symmetric, which implies that this situation is unlikely to generate enough flux variability. Rotating hotspots are more likely to generate QPOs if the observer's line of sight is close to the plane of the accretion disk \citep{2009ApJ...696.2170R}.

Another interesting model that might explain fast quasi-periodicity in the jet emission involves magnetic reconnection events in almost equidistant magnetic islands inside the jet \citep{Huang_2013}. These equispaced magnetic islands periodically enhance the flux, thereby producing a rapid transient QPO. \citet{Shukla_2018} used such a magnetic reconnection process to model the extremely fast variability ($\sim$5 minutes) in the FSRQ CTA 102 during its outburst in 2016. This model does appear to be capable of producing the 3.6-day QPO in $\gamma$-rays for PKS 1510-089, and suffers no obvious difficulties.

We can also attempt to attribute both the 3.6-day and 92-day QPO of PKS 1510-089 to a very reasonable model that involves a plasma blob moving helically down the jet \citep{Mohan_2015, 10.1093/mnras/stw2684, 10.1093/mnras/staa3211}. For the simplest leptonic one-zone model (where the bulk of the synchrotron and inverse-Compton emission comes from a single region), such a plasma blob contains higher particle and magnetic energy densities and is responsible for occasional enhanced emission from blazars. Due to the postulated helical motion of the blob, the viewing angle of the blob with respect to our line of sight ($\theta_{\text{obs}}$) changes periodically with time as,
\begin{equation}
    \cos\theta_{\text{obs}}(t) = \sin\phi\sin\psi\cos(2\upi t/P_{\text{obs}})+\cos\phi\cos\psi ,
\end{equation}
where $\phi$ is the pitch angle of the helical path, $\psi$ is the angle of the jet axis with respect to our line of sight, and $P_{\text{obs}}$ is the observed periodicity in the light curve \citep{10.1093/mnras/stw2684, 2018NatCo...9.4599Z}. The Doppler factor ($\delta$) varies with the viewing angle as $\delta = 1/[\Gamma(1-\beta\cos\theta_{\text{obs}})]$, where $\Gamma = 1/\sqrt{1-\beta^2}$ is the bulk Lorentz factor of the blob motion with $\beta = v_{\text{jet}}/c$. Then the periodicity in the blob rest frame is given as,
\begin{equation}
    P_{\text{rf}} = \frac{P_{\text{obs}}}{1-\beta\cos\psi\cos\phi} .
\end{equation}
This model can naturally explain the transient nature of any periodicities as the QPO starts when the blob is injected into the jet and lasts until the blob dissipates. One limitation of this model is that it can only explain a QPO having almost constant amplitude.  It is evident from \autoref{fig:wwzep1} and \autoref{fig:wwzep2} that both the QPOs in EP1 and EP2 have varying amplitudes.  We note that \citet{10.1093/mnras/staa3211} used a curved jet scenario to model a transient QPO with varying amplitude in 3C 454.3. In this model, the angle between the jet axis and our line of sight ($\psi$) becomes time-dependent ($\psi(t)$). We find that the "blob moving helically in a curved jet" model might be able to  explain them both. Assuming $\phi \simeq 2^{\circ}$ \citep{2018NatCo...9.4599Z}, $\langle \psi \rangle = 2^{\circ}.2$, $\Gamma=20.0$ \citep{10.1093/mnras/stab975} and $P_{\text{obs}}=3.6$ days, the periodicity in the blob rest-frame is $P_{\text{rf}} \simeq 3.8$ years. The blob traverses about a distance $D = c\beta P_{\text{rf}} \cos\phi \simeq 1.16$ parsec down the jet during one period. But only a very high jet curvature can explain the rapid changes in QPO amplitudes in EP1. However, the presence of high curvature within a few parsecs is highly unlikely in an extremely long, well-collimated, and powerful FSRQ jet. On the other hand, for $P_{\text{obs}}=92$ days, the periodicity in the blob rest-frame is $P_{\text{rf}} \simeq 97.1$ years. The blob travels $\sim$30 pc during one period, i.e., it travels $\sim$200 pc during EP2. \autoref{fig:wwzep2} shows a slow increasing trend in the first 5 cycles and then a faster attenuation in the QPO amplitudes. This situation can be explained with a much lower curvature in the jet, which is physically more likely. Thus, we tentatively conclude that the 3.6-day QPO  probably resulted from flux enhancements by magnetic reconnection events at quasi-equispaced magnetic islands, while most probably the origin of the 92-day QPO is a plasma blob  moving helically inside a curved jet.

\section{Conclusions}
\label{sec:con}

In this work, we report two transient QPOs in $\gamma$-rays exhibited by the TeV blazar PKS 1510-089. We employed several standard tools such as Lomb-Scargle periodogram, \texttt{REDFIT}, WWZ and lightcurve simulation on the \textit{Fermi}-LAT light curve to detect the significant periodicities. Our key results are as follows.

\begin{enumerate}
    \item PKS 1510-089 showed a fast periodicity of $\sim3.6$ days during a flare in 2009, from MJD 54906 to MJD 54923. It lasted for only 18 days. Lightcurve simulation indicates the significance of this QPO to be $\sim3.5\sigma$ against the underlying red-noise spectrum. To our knowledge, this is the shortest period  QPO so far reported in the $\gamma$-ray emission of a blazar.
    
    \item Multiple QPO analyses show the presence of a QPO around 92 days in the recent \textit{Fermi}-LAT observations that lasted for about 650 days from 2018 to 2020 (MJD 58200--58850). From light curve simulation, it seems that this periodicity has a significance of about 7$\sigma$, and would be the most significant blazar QPO ever reported.
\end{enumerate}

\section*{Acknowledgements}

This research has made use of the Fermi-LAT data, obtained from the Fermi Science Support Center, provided by NASA’s Goddard Space Flight Center (GSFC). The {\it Fermi} LAT Collaboration acknowledges generous on-going support from a number of agencies and institutes that
have  supported  both  the  development and  the  operation  of
the LAT as well as scientific data analysis. These include the
National Aeronautics and Space Administration and the Department of Energy in the United States, the Commissariat 
l' Energie Atomique and the Centre National de la Recherche
Scientifique /  Institut National de  Physique Nucl\'eaire et  de
Physique des Particules in France, the Agenzia Spaziale Italiana and the Istituto Nazionale di Fisica Nucleare in Italy,
the
Ministry of Education, Culture, Sports, Science and Technology (MEXT), High Energy Accelerator Research Organization (KEK) and Japan Aerospace Exploration Agency (JAXA)
in Japan, and the K\. A.\ Wallenberg Foundation, the Swedish
Research Council and the Swedish National Space Board in
Sweden.  Additional support for science analysis during the
operations  phase  is  gratefully  acknowledged  from  the  Istituto Nazionale di Astrofisica in Italy and the Centre National  d'  \'Etudes Spatiales in France.The data and analysis software were obtained from NASA’s High-Energy Astrophysics Science Archive Research Center (HEASARC), a service of GSFC. We used a community-developed Python package named \texttt{ENRICO} to make Fermi-LAT data analysis easier   \citep{sanchez2013enrico}. Finally, We acknowledge the support of the Department of Atomic Energy, Government of India, under Project Identification No. RTI 4002.

%%%%%%%%%%%%%%%%%%%%%%%%%%%%%%%%%%%%%%%%%%%%%%%%%%
\section*{Data Availability}

\begin{enumerate}
    \item The Fermi-LAT data used in this article are available in the
LAT data server at \url{https://fermi.gsfc.nasa.gov/ssc/data/access/}.
    \item The Fermi-LAT data analysis software is available at \url{https://fermi.gsfc.nasa.gov/ssc/data/analysis/software/}.
    \item We agree to share data derived in this article on reasonable request to the corresponding author.

\end{enumerate}



\bibliographystyle{mnras}
\bibliography{example} % if your bibtex file is called example.bib

\begin{thebibliography}{}
\makeatletter
\relax
\def\mn@urlcharsother{\let\do\@makeother \do\$\do\&\do\#\do\^\do\_\do\%\do\~}
\def\mn@doi{\begingroup\mn@urlcharsother \@ifnextchar [ {\mn@doi@}
  {\mn@doi@[]}}
\def\mn@doi@[#1]#2{\def\@tempa{#1}\ifx\@tempa\@empty \href
  {http://dx.doi.org/#2} {doi:#2}\else \href {http://dx.doi.org/#2} {#1}\fi
  \endgroup}
\def\mn@eprint#1#2{\mn@eprint@#1:#2::\@nil}
\def\mn@eprint@arXiv#1{\href {http://arxiv.org/abs/#1} {{\tt arXiv:#1}}}
\def\mn@eprint@dblp#1{\href {http://dblp.uni-trier.de/rec/bibtex/#1.xml}
  {dblp:#1}}
\def\mn@eprint@#1:#2:#3:#4\@nil{\def\@tempa {#1}\def\@tempb {#2}\def\@tempc
  {#3}\ifx \@tempc \@empty \let \@tempc \@tempb \let \@tempb \@tempa \fi \ifx
  \@tempb \@empty \def\@tempb {arXiv}\fi \@ifundefined
  {mn@eprint@\@tempb}{\@tempb:\@tempc}{\expandafter \expandafter \csname
  mn@eprint@\@tempb\endcsname \expandafter{\@tempc}}}

\bibitem[\protect\citeauthoryear{Abdollahi et~al.,}{Abdollahi
  et~al.}{2020}]{Abdollahi_2020}
Abdollahi S.,  et~al., 2020, \mn@doi [\apjs] {10.3847/1538-4365/ab6bcb}, 247,
  33

\bibitem[\protect\citeauthoryear{Ackermann et~al.,}{Ackermann
  et~al.}{2015}]{Ackermann_2015}
Ackermann M.,  et~al., 2015, \mn@doi [\apj] {10.1088/2041-8205/813/2/l41}, 813,
  L41

\bibitem[\protect\citeauthoryear{Aleksi{\'{c}} et~al.,}{Aleksi{\'{c}}
  et~al.}{2011}]{Aleksi__2011}
Aleksi{\'{c}} J.,  et~al., 2011, \mn@doi [\apj] {10.1088/2041-8205/730/1/l8},
  730, L8

\bibitem[\protect\citeauthoryear{{Atwood} et~al.,}{{Atwood}
  et~al.}{2009}]{2009ApJ...697.1071A}
{Atwood} W.~B.,  et~al., 2009, \mn@doi [\apj] {10.1088/0004-637X/697/2/1071},
  \href {https://ui.adsabs.harvard.edu/abs/2009ApJ...697.1071A} {697, 1071}

\bibitem[\protect\citeauthoryear{{Barnacka}, {Moderski}, {Behera}, {Brun}  \&
  {Wagner}}{{Barnacka} et~al.}{2014}]{2014A&A...567A.113B}
{Barnacka} A.,  {Moderski} R.,  {Behera} B.,  {Brun} P.,   {Wagner} S.,  2014,
  \mn@doi [\aap] {10.1051/0004-6361/201322205}, \href
  {https://ui.adsabs.harvard.edu/abs/2014A&A...567A.113B} {567, A113}

\bibitem[\protect\citeauthoryear{{Beaklini}, {Dominici}  \&
  {Abraham}}{{Beaklini} et~al.}{2017}]{2017A&A...606A..87B}
{Beaklini} P. P.~B.,  {Dominici} T.~P.,   {Abraham} Z.,  2017, \mn@doi [\aap]
  {10.1051/0004-6361/201731118}, \href
  {https://ui.adsabs.harvard.edu/abs/2017A&A...606A..87B} {606, A87}

\bibitem[\protect\citeauthoryear{Bhatta}{Bhatta}{2019}]{10.1093/mnras/stz1482}
Bhatta G.,  2019, \mn@doi [\mnras] {10.1093/mnras/stz1482}, 487, 3990

\bibitem[\protect\citeauthoryear{{Castignani} et~al.,}{{Castignani}
  et~al.}{2017}]{2017A&A...601A..30C}
{Castignani} G.,  et~al., 2017, \mn@doi [\aap] {10.1051/0004-6361/201629775},
  \href {https://ui.adsabs.harvard.edu/abs/2017A&A...601A..30C} {601, A30}

\bibitem[\protect\citeauthoryear{{Connolly}}{{Connolly}}{2016}]{2016ascl.soft02012C}
{Connolly} S.~D.,  2016, {DELightcurveSimulation: Light curve simulation code}
  (\mn@eprint {ascl} {1602.012})

\bibitem[\protect\citeauthoryear{Covino, Sandrinelli  \& Treves}{Covino
  et~al.}{2018}]{10.1093/mnras/sty2720}
Covino S.,  Sandrinelli A.,   Treves A.,  2018, \mn@doi [\mnras]
  {10.1093/mnras/sty2720}, 482, 1270

\bibitem[\protect\citeauthoryear{{Covino}, {Sandrinelli}  \& {Treves}}{{Covino}
  et~al.}{2019}]{2019MNRAS.482.1270C}
{Covino} S.,  {Sandrinelli} A.,   {Treves} A.,  2019, \mn@doi [\mnras]
  {10.1093/mnras/sty2720}, \href
  {https://ui.adsabs.harvard.edu/abs/2019MNRAS.482.1270C} {482, 1270}

\bibitem[\protect\citeauthoryear{{Czesla}, {Schr{\"o}ter}, {Schneider},
  {Huber}, {Pfeifer}, {Andreasen}  \& {Zechmeister}}{{Czesla}
  et~al.}{2019}]{pya}
{Czesla} S.,  {Schr{\"o}ter} S.,  {Schneider} C.~P.,  {Huber} K.~F.,  {Pfeifer}
  F.,  {Andreasen} D.~T.,   {Zechmeister} M.,  2019, {PyA: Python
  astronomy-related packages} (\mn@eprint {ascl} {1906.010})

\bibitem[\protect\citeauthoryear{{Emmanoulopoulos}, {McHardy}  \&
  {Papadakis}}{{Emmanoulopoulos} et~al.}{2013}]{2013MNRAS.433..907E}
{Emmanoulopoulos} D.,  {McHardy} I.~M.,   {Papadakis} I.~E.,  2013, \mn@doi
  [\mnras] {10.1093/mnras/stt764}, \href
  {https://ui.adsabs.harvard.edu/abs/2013MNRAS.433..907E} {433, 907}

\bibitem[\protect\citeauthoryear{{Fan} et~al.,}{{Fan}
  et~al.}{2007}]{2007A&A...462..547F}
{Fan} J.~H.,  et~al., 2007, \mn@doi [\aap] {10.1051/0004-6361:20054775}, \href
  {https://ui.adsabs.harvard.edu/abs/2007A&A...462..547F} {462, 547}

\bibitem[\protect\citeauthoryear{{Fermi Science Support Development
  Team}}{{Fermi Science Support Development Team}}{2019}]{2019ascl.soft05011F}
{Fermi Science Support Development Team} 2019, {Fermitools: Fermi Science
  Tools} (\mn@eprint {ascl} {1905.011})

\bibitem[\protect\citeauthoryear{{Foster}}{{Foster}}{1996}]{1996AJ....112.1709F}
{Foster} G.,  1996, \mn@doi [\aj] {10.1086/118137}, \href
  {https://ui.adsabs.harvard.edu/abs/1996AJ....112.1709F} {112, 1709}

\bibitem[\protect\citeauthoryear{{Fu}, {Zhang}, {Zhang}, {Xiong}  \&
  {Guo}}{{Fu} et~al.}{2014}]{2014AcASn..55....1F}
{Fu} J.~P.,  {Zhang} H.~J.,  {Zhang} X.,  {Xiong} D.~R.,   {Guo} F.,  2014,
  AcASn, \href {https://ui.adsabs.harvard.edu/abs/2014AcASn..55....1F} {55, 1}

\bibitem[\protect\citeauthoryear{Grossmann \& Morlet}{Grossmann \&
  Morlet}{1984}]{doi:10.1137/0515056}
Grossmann A.,  Morlet J.,  1984, \mn@doi [SJMA] {10.1137/0515056}, 15, 723

\bibitem[\protect\citeauthoryear{Gu, Cao  \& Jiang}{Gu
  et~al.}{2001}]{10.1046/j.1365-8711.2001.04795.x}
Gu M.,  Cao X.,   Jiang D.,  2001, \mn@doi [\mnras]
  {10.1046/j.1365-8711.2001.04795.x}, 327, 1111

\bibitem[\protect\citeauthoryear{{Gupta}}{{Gupta}}{2014}]{2014JApA...35..307G}
{Gupta} A.~C.,  2014, \mn@doi [JApA] {10.1007/s12036-014-9219-7}, \href
  {https://ui.adsabs.harvard.edu/abs/2014JApA...35..307G} {35, 307}

\bibitem[\protect\citeauthoryear{{Gupta}}{{Gupta}}{2018}]{2018Galax...6....1G}
{Gupta} A.,  2018, \mn@doi [Galaxies] {10.3390/galaxies6010001}, \href
  {https://ui.adsabs.harvard.edu/abs/2018Galax...6....1G} {6, 1}

\bibitem[\protect\citeauthoryear{{Gupta}, {Srivastava}  \& {Wiita}}{{Gupta}
  et~al.}{2009}]{2009ApJ...690..216G}
{Gupta} A.~C.,  {Srivastava} A.~K.,   {Wiita} P.~J.,  2009, \mn@doi [\apj]
  {10.1088/0004-637X/690/1/216}, \href
  {https://ui.adsabs.harvard.edu/abs/2009ApJ...690..216G} {690, 216}

\bibitem[\protect\citeauthoryear{{Gupta}, {Tripathi}, {Wiita}, {Kushwaha},
  {Zhang}  \& {Bambi}}{{Gupta} et~al.}{2019}]{2019MNRAS.484.5785G}
{Gupta} A.~C.,  {Tripathi} A.,  {Wiita} P.~J.,  {Kushwaha} P.,  {Zhang} Z.,
  {Bambi} C.,  2019, \mn@doi [\mnras] {10.1093/mnras/stz395}, \href
  {https://ui.adsabs.harvard.edu/abs/2019MNRAS.484.5785G} {484, 5785}

\bibitem[\protect\citeauthoryear{{H.~E.~S.~S. Collaboration}
  et~al.,}{{H.~E.~S.~S. Collaboration} et~al.}{2013}]{2013A&A...554A.107H}
{H.~E.~S.~S. Collaboration} et~al., 2013, \mn@doi [\aap]
  {10.1051/0004-6361/201321135}, \href
  {https://ui.adsabs.harvard.edu/abs/2013A&A...554A.107H} {554, A107}

\bibitem[\protect\citeauthoryear{Huang, Wang, Wang  \& Wang}{Huang
  et~al.}{2013}]{Huang_2013}
Huang C.-Y.,  Wang D.-X.,  Wang J.-Z.,   Wang Z.-Y.,  2013, \mn@doi [RAA]
  {10.1088/1674-4527/13/6/010}, 13, 705

\bibitem[\protect\citeauthoryear{King et~al.,}{King
  et~al.}{2013}]{10.1093/mnrasl/slt125}
King O.~G.,  et~al., 2013, \mn@doi [\mnras] {10.1093/mnrasl/slt125}, 436, L114

\bibitem[\protect\citeauthoryear{Kushwaha, Sinha, Misra, Singh  \& de Gouveia
  Dal~Pino}{Kushwaha et~al.}{2017}]{Kushwaha_2017}
Kushwaha P.,  Sinha A.,  Misra R.,  Singh K.~P.,   de Gouveia Dal~Pino E.~M.,
  2017, \mn@doi [\apj] {10.3847/1538-4357/aa8ef5}, 849, 138

\bibitem[\protect\citeauthoryear{{Kushwaha}, {Sarkar}, {Gupta}, {Tripathi}  \&
  {Wiita}}{{Kushwaha} et~al.}{2020a}]{2020MNRAS.499..653K}
{Kushwaha} P.,  {Sarkar} A.,  {Gupta} A.~C.,  {Tripathi} A.,   {Wiita} P.~J.,
  2020a, \mn@doi [\mnras] {10.1093/mnras/staa2899}, \href
  {https://ui.adsabs.harvard.edu/abs/2020MNRAS.499..653K} {499, 653}

\bibitem[\protect\citeauthoryear{Kushwaha, Sarkar, Gupta, Tripathi  \&
  Wiita}{Kushwaha et~al.}{2020b}]{10.1093/mnras/staa2899}
Kushwaha P.,  Sarkar A.,  Gupta A.~C.,  Tripathi A.,   Wiita P.~J.,  2020b,
  \mn@doi [\mnras] {10.1093/mnras/staa2899}, 499, 653

\bibitem[\protect\citeauthoryear{{Lachowicz}, {Gupta}, {Gaur}  \&
  {Wiita}}{{Lachowicz} et~al.}{2009}]{2009A&A...506L..17L}
{Lachowicz} P.,  {Gupta} A.~C.,  {Gaur} H.,   {Wiita} P.~J.,  2009, \mn@doi
  [\aap] {10.1051/0004-6361/200913161}, \href
  {https://ui.adsabs.harvard.edu/abs/2009A&A...506L..17L} {506, L17}

\bibitem[\protect\citeauthoryear{{Li}, {Fan}  \& {Yuan}}{{Li}
  et~al.}{2007}]{2007ChPhy..16..876L}
{Li} J.,  {Fan} J.-H.,   {Yuan} Y.-H.,  2007, \mn@doi [Chinese Physics]
  {10.1088/1009-1963/16/3/053}, \href
  {https://ui.adsabs.harvard.edu/abs/2007ChPhy..16..876L} {16, 876}

\bibitem[\protect\citeauthoryear{{Li}, {Zhao}, {Yan}, {Wang}, {Yang}, {Cai}  \&
  {Luo}}{{Li} et~al.}{2021}]{2021JApA...42...92L}
{Li} X.-P.,  {Zhao} L.,  {Yan} Y.,  {Wang} L.-S.,  {Yang} H.-T.,  {Cai} Y.,
  {Luo} Y.-H.,  2021, \mn@doi [JApA] {10.1007/s12036-021-09773-9}, \href
  {https://ui.adsabs.harvard.edu/abs/2021JApA...42...92L} {42, 92}

\bibitem[\protect\citeauthoryear{{Liska}, {Hesp}, {Tchekhovskoy}, {Ingram},
  {van der Klis}  \& {Markoff}}{{Liska} et~al.}{2018}]{2018MNRAS.474L..81L}
{Liska} M.,  {Hesp} C.,  {Tchekhovskoy} A.,  {Ingram} A.,  {van der Klis} M.,
  {Markoff} S.,  2018, \mn@doi [\mnras] {10.1093/mnrasl/slx174}, \href
  {https://ui.adsabs.harvard.edu/abs/2018MNRAS.474L..81L} {474, L81}

\bibitem[\protect\citeauthoryear{Liu, Zhao  \& Wu}{Liu et~al.}{2006}]{Liu_2006}
Liu F.~K.,  Zhao G.,   Wu X.-B.,  2006, \mn@doi [\apj] {10.1086/507267}, 650,
  749

\bibitem[\protect\citeauthoryear{{Lomb}}{{Lomb}}{1976}]{1976Ap&SS..39..447L}
{Lomb} N.~R.,  1976, \mn@doi [\apss] {10.1007/BF00648343}, \href
  {https://ui.adsabs.harvard.edu/abs/1976Ap&SS..39..447L} {39, 447}

\bibitem[\protect\citeauthoryear{Meyer, Scargle  \& Blandford}{Meyer
  et~al.}{2019}]{Meyer_2019}
Meyer M.,  Scargle J.~D.,   Blandford R.~D.,  2019, \mn@doi [\apj]
  {10.3847/1538-4357/ab1651}, 877, 39

\bibitem[\protect\citeauthoryear{Mohan \& Mangalam}{Mohan \&
  Mangalam}{2015}]{Mohan_2015}
Mohan P.,  Mangalam A.,  2015, \mn@doi [\apj] {10.1088/0004-637x/805/2/91},
  805, 91

\bibitem[\protect\citeauthoryear{{Patel}, {Bose}, {Gupta}  \& {Zuberi}}{{Patel}
  et~al.}{2021}]{2021JHEAp..29...31P}
{Patel} S.~R.,  {Bose} D.,  {Gupta} N.,   {Zuberi} M.,  2021, \mn@doi [JHEAp]
  {10.1016/j.jheap.2020.12.001}, \href
  {https://ui.adsabs.harvard.edu/abs/2021JHEAp..29...31P} {29, 31}

\bibitem[\protect\citeauthoryear{{Pe{\~n}il} et~al.,}{{Pe{\~n}il}
  et~al.}{2020}]{2020ApJ...896..134P}
{Pe{\~n}il} P.,  et~al., 2020, \mn@doi [\apj] {10.3847/1538-4357/ab910d}, \href
  {https://ui.adsabs.harvard.edu/abs/2020ApJ...896..134P} {896, 134}

\bibitem[\protect\citeauthoryear{Percival \& Walden}{Percival \&
  Walden}{1993}]{percival_walden_1993}
Percival D.~B.,  Walden A.~T.,  1993, Spectral Analysis for Physical
  Applications.
Cambridge University Press, \mn@doi{10.1017/CBO9780511622762}

\bibitem[\protect\citeauthoryear{Prince, Majumdar  \& Gupta}{Prince
  et~al.}{2017}]{Prince_2017}
Prince R.,  Majumdar P.,   Gupta N.,  2017, \mn@doi [\apj]
  {10.3847/1538-4357/aa78f4}, 844, 62

\bibitem[\protect\citeauthoryear{{Raiteri} et~al.,}{{Raiteri}
  et~al.}{2001}]{2001A&A...377..396R}
{Raiteri} C.~M.,  et~al., 2001, \mn@doi [\aap] {10.1051/0004-6361:20011112},
  \href {https://ui.adsabs.harvard.edu/abs/2001A&A...377..396R} {377, 396}

\bibitem[\protect\citeauthoryear{{Rakshit}}{{Rakshit}}{2020}]{2020A&A...642A..59R}
{Rakshit} S.,  2020, \mn@doi [\aap] {10.1051/0004-6361/202038324}, \href
  {https://ui.adsabs.harvard.edu/abs/2020A&A...642A..59R} {642, A59}

\bibitem[\protect\citeauthoryear{{Rani}, {Wiita}  \& {Gupta}}{{Rani}
  et~al.}{2009}]{2009ApJ...696.2170R}
{Rani} B.,  {Wiita} P.~J.,   {Gupta} A.~C.,  2009, \mn@doi [\apj]
  {10.1088/0004-637X/696/2/2170}, \href
  {https://ui.adsabs.harvard.edu/abs/2009ApJ...696.2170R} {696, 2170}

\bibitem[\protect\citeauthoryear{Rieger}{Rieger}{2004}]{Rieger_2004}
Rieger F.~M.,  2004, \mn@doi [\apj] {10.1086/426018}, 615, L5

\bibitem[\protect\citeauthoryear{Robinson}{Robinson}{1977}]{RePEc:eee:spapps:v:6:y:1977:i:1:p:9-24}
Robinson P.~M.,  1977, Stochastic Processes and their Applications, 6, 9

\bibitem[\protect\citeauthoryear{{Romero}, {Chajet}, {Abraham}  \&
  {Fan}}{{Romero} et~al.}{2000}]{2000A&A...360...57R}
{Romero} G.~E.,  {Chajet} L.,  {Abraham} Z.,   {Fan} J.~H.,  2000, \aap, \href
  {https://ui.adsabs.harvard.edu/abs/2000A&A...360...57R} {360, 57}

\bibitem[\protect\citeauthoryear{Roy, Patel, Sarkar, Chatterjee  \&
  Chitnis}{Roy et~al.}{2021}]{10.1093/mnras/stab975}
Roy A.,  Patel S.~R.,  Sarkar A.,  Chatterjee A.,   Chitnis V.~R.,  2021,
  \mn@doi [\mnras] {10.1093/mnras/stab975}, 504, 1103

\bibitem[\protect\citeauthoryear{Sanchez \& Deil}{Sanchez \&
  Deil}{2013}]{sanchez2013enrico}
Sanchez D.~A.,  Deil C.,  2013, Enrico : a Python package to simplify Fermi-LAT
  analysis (\mn@eprint {arXiv} {1307.4534})

\bibitem[\protect\citeauthoryear{Sandrinelli, Covino, Dotti  \&
  Treves}{Sandrinelli et~al.}{2016a}]{Sandrinelli_2016}
Sandrinelli A.,  Covino S.,  Dotti M.,   Treves A.,  2016a, \mn@doi [\aj]
  {10.3847/0004-6256/151/3/54}, 151, 54

\bibitem[\protect\citeauthoryear{Sandrinelli, Covino  \& Treves}{Sandrinelli
  et~al.}{2016b}]{Sandrinelli_2016a}
Sandrinelli A.,  Covino S.,   Treves A.,  2016b, \mn@doi [\apj]
  {10.3847/0004-637x/820/1/20}, 820, 20

\bibitem[\protect\citeauthoryear{{Sandrinelli} et~al.,}{{Sandrinelli}
  et~al.}{2017}]{2017A&A...600A.132S}
{Sandrinelli} A.,  et~al., 2017, \mn@doi [\aap] {10.1051/0004-6361/201630288},
  \href {https://ui.adsabs.harvard.edu/abs/2017A&A...600A.132S} {600, A132}

\bibitem[\protect\citeauthoryear{Sarkar, Gupta, Chitnis  \& Wiita}{Sarkar
  et~al.}{2020a}]{10.1093/mnras/staa3211}
Sarkar A.,  Gupta A.~C.,  Chitnis V.~R.,   Wiita P.~J.,  2020a, \mn@doi
  [\mnras] {10.1093/mnras/staa3211}, 501, 50

\bibitem[\protect\citeauthoryear{{Sarkar}, {Kushwaha}, {Gupta}, {Chitnis}  \&
  {Wiita}}{{Sarkar} et~al.}{2020b}]{2020A&A...642A.129S}
{Sarkar} A.,  {Kushwaha} P.,  {Gupta} A.~C.,  {Chitnis} V.~R.,   {Wiita} P.~J.,
   2020b, \mn@doi [\aap] {10.1051/0004-6361/202038052}, \href
  {https://ui.adsabs.harvard.edu/abs/2020A&A...642A.129S} {642, A129}

\bibitem[\protect\citeauthoryear{{Scargle}}{{Scargle}}{1982}]{1982ApJ...263..835S}
{Scargle} J.~D.,  1982, \mn@doi [\apj] {10.1086/160554}, \href
  {https://ui.adsabs.harvard.edu/abs/1982ApJ...263..835S} {263, 835}

\bibitem[\protect\citeauthoryear{Schulz \& Mudelsee}{Schulz \&
  Mudelsee}{2002}]{SCHULZ2002421}
Schulz M.,  Mudelsee M.,  2002, \mn@doi [Computers & Geosciences]
  {https://doi.org/10.1016/S0098-3004(01)00044-9}, 28, 421

\bibitem[\protect\citeauthoryear{Shukla et~al.,}{Shukla
  et~al.}{2018}]{Shukla_2018}
Shukla A.,  et~al., 2018, \mn@doi [\apj] {10.3847/2041-8213/aaacca}, 854, L26

\bibitem[\protect\citeauthoryear{Sobacchi, Sormani  \& Stamerra}{Sobacchi
  et~al.}{2016}]{10.1093/mnras/stw2684}
Sobacchi E.,  Sormani M.~C.,   Stamerra A.,  2016, \mn@doi [\mnras]
  {10.1093/mnras/stw2684}, 465, 161

\bibitem[\protect\citeauthoryear{{Stella} \& {Vietri}}{{Stella} \&
  {Vietri}}{1998}]{1998ApJ...492L..59S}
{Stella} L.,  {Vietri} M.,  1998, \mn@doi [\apjl] {10.1086/311075}, \href
  {https://ui.adsabs.harvard.edu/abs/1998ApJ...492L..59S} {492, L59}

\bibitem[\protect\citeauthoryear{Urry \& Padovani}{Urry \&
  Padovani}{1995}]{Urry_1995}
Urry C.~M.,  Padovani P.,  1995, \mn@doi [\pasp] {10.1086/133630}, 107, 803

\bibitem[\protect\citeauthoryear{{Valtonen} et~al.,}{{Valtonen}
  et~al.}{2008}]{2008Natur.452..851V}
{Valtonen} M.~J.,  et~al., 2008, \mn@doi [\nat] {10.1038/nature06896}, \href
  {https://ui.adsabs.harvard.edu/abs/2008Natur.452..851V} {452, 851}

\bibitem[\protect\citeauthoryear{{Vaughan}}{{Vaughan}}{2005}]{2005A&A...431..391V}
{Vaughan} S.,  2005, \mn@doi [\aap] {10.1051/0004-6361:20041453}, \href
  {https://ui.adsabs.harvard.edu/abs/2005A&A...431..391V} {431, 391}

\bibitem[\protect\citeauthoryear{Villforth et~al.,}{Villforth
  et~al.}{2010}]{10.1111/j.1365-2966.2009.16133.x}
Villforth C.,  et~al., 2010, \mn@doi [\mnras]
  {10.1111/j.1365-2966.2009.16133.x}, 402, 2087

\bibitem[\protect\citeauthoryear{Welch}{Welch}{1967}]{1161901}
Welch P.,  1967, \mn@doi [IEEE Transactions on Audio and Electroacoustics]
  {10.1109/TAU.1967.1161901}, 15, 70

\bibitem[\protect\citeauthoryear{Wu, Zhou, Peng, Ma, Jiang  \& Chen}{Wu
  et~al.}{2005}]{10.1111/j.1365-2966.2005.09150.x}
Wu J.,  Zhou X.,  Peng B.,  Ma J.,  Jiang Z.,   Chen J.,  2005, \mn@doi
  [\mnras] {10.1111/j.1365-2966.2005.09150.x}, 361, 155

\bibitem[\protect\citeauthoryear{{Xie}, {Liu}, {Cha}, {Zhou}, {Ma}, {Xie}  \&
  {Chen}}{{Xie} et~al.}{2005}]{2005AJ....130.2506X}
{Xie} G.~Z.,  {Liu} H.~T.,  {Cha} G.~W.,  {Zhou} S.~B.,  {Ma} L.,  {Xie} Z.~H.,
    {Chen} L.~E.,  2005, \mn@doi [\aj] {10.1086/497163}, \href
  {https://ui.adsabs.harvard.edu/abs/2005AJ....130.2506X} {130, 2506}

\bibitem[\protect\citeauthoryear{{Xie}, {Yi}, {Li}, {Zhou}  \& {Chen}}{{Xie}
  et~al.}{2008}]{2008AJ....135.2212X}
{Xie} G.~Z.,  {Yi} T.~F.,  {Li} H.~Z.,  {Zhou} S.~B.,   {Chen} L.~E.,  2008,
  \mn@doi [\aj] {10.1088/0004-6256/135/6/2212}, \href
  {https://ui.adsabs.harvard.edu/abs/2008AJ....135.2212X} {135, 2212}

\bibitem[\protect\citeauthoryear{{Zacharias}}{{Zacharias}}{2018}]{2018heas.confE..33Z}
{Zacharias} M.,  2018, in High Energy Astrophysics in Southern Africa
  (HEASA2018). p.~33 (\mn@eprint {arXiv} {1903.02274})

\bibitem[\protect\citeauthoryear{{Zechmeister} \& {K{\"u}rster}}{{Zechmeister}
  \& {K{\"u}rster}}{2009}]{2009A&A...496..577Z}
{Zechmeister} M.,  {K{\"u}rster} M.,  2009, \mn@doi [\aap]
  {10.1051/0004-6361:200811296}, \href
  {https://ui.adsabs.harvard.edu/abs/2009A&A...496..577Z} {496, 577}

\bibitem[\protect\citeauthoryear{{Zhang} \& {Bao}}{{Zhang} \&
  {Bao}}{1991}]{1991A&A...246...21Z}
{Zhang} X.~H.,  {Bao} G.,  1991, \aap, \href
  {https://ui.adsabs.harvard.edu/abs/1991A&A...246...21Z} {246, 21}

\bibitem[\protect\citeauthoryear{{Zhang}, {Zhao}, {Zhang}, {Dong}, {Xie}, {Yi},
  {Zheng}  \& {Yu}}{{Zhang} et~al.}{2009}]{2009ScChG..52.1442Z}
{Zhang} H.,  {Zhao} G.,  {Zhang} X.,  {Dong} F.,  {Xie} Z.,  {Yi} T.,  {Zheng}
  Y.,   {Yu} Y.,  2009, \mn@doi [ScChG] {10.1007/s11433-009-0175-1}, \href
  {https://ui.adsabs.harvard.edu/abs/2009ScChG..52.1442Z} {52, 1442}

\bibitem[\protect\citeauthoryear{{Zhou}, {Wang}, {Chen}, {Wiita},
  {Vadakkumthani}, {Morrell}, {Zhang}  \& {Zhang}}{{Zhou}
  et~al.}{2018}]{2018NatCo...9.4599Z}
{Zhou} J.,  {Wang} Z.,  {Chen} L.,  {Wiita} P.~J.,  {Vadakkumthani} J.,
  {Morrell} N.,  {Zhang} P.,   {Zhang} J.,  2018, \mn@doi [NatCo]
  {10.1038/s41467-018-07103-2}, \href
  {https://ui.adsabs.harvard.edu/abs/2018NatCo...9.4599Z} {9, 4599}

\makeatother
\end{thebibliography}

%\begin{thebibliography}{99}
%\end{thebibliography}

%%%%%%%%%%%%%%%%%%%%%%%%%%%%%%%%%%%%%%%%%%%%%%%%%%

%%%%%%%%%%%%%%%%% APPENDICES %%%%%%%%%%%%%%%%%%%%%

%\appendix


%it can be placed in an Appendix, which appears after the list of references.

%%%%%%%%%%%%%%%%%%%%%%%%%%%%%%%%%%%%%%%%%%%%%%%%%%

% Don't change these lines
\bsp	% typesetting comment
\label{lastpage}
\end{document}